\documentclass[%
 aip,
 amsmath,amssymb,
 reprint,%
]{revtex4-1}

\usepackage{graphicx}
\usepackage{dcolumn}
\usepackage{bm}

\usepackage[utf8]{inputenc}
\usepackage[T1]{fontenc}
\usepackage{mathptmx}
\usepackage{etoolbox}
\usepackage{color}

\makeatletter
\def\@email#1#2{%
 \endgroup
 \patchcmd{\titleblock@produce}
  {\frontmatter@RRAPformat}
  {\frontmatter@RRAPformat{\produce@RRAP{*#1\href{mailto:#2}{#2}}}\frontmatter@RRAPformat}
  {}{}
}%
\makeatother
\begin{document}

\preprint{AIP/123-QED}

\title[Topological Data Analysis of Ion Migration Mechanism]{Topological Data Analysis of Ion Migration Mechanism}
\author{Ryuhei Sato}%
\affiliation{Advanced Institute for Materials Research, Tohoku University, 2-1-1 Katahira, Aoba-ku, Sendai 980-8577, Japan}

\author{Kazuto Akagi}%
\affiliation{Advanced Institute for Materials Research, Tohoku University, 2-1-1 Katahira, Aoba-ku, Sendai 980-8577, Japan}

\author{Shigeyuki Takagi}%
\affiliation{Institute for Materials Research, Tohoku University, 2-1-1 Katahira, Aoba-ku, Sendai 980-8577, Japan.}

\author{Kartik Sau}%
\affiliation{Advanced Institute for Materials Research, Tohoku University, 2-1-1 Katahira, Aoba-ku, Sendai 980-8577, Japan}
\affiliation{Mathematics for Advanced Materials Open Innovation Laboratory (MathAM-OIL), National Institute of Advanced Industrial Science and Technology (AIST), c/o Advanced Institute for Materials Research (AIMR), Tohoku University, Sendai 980-8577, Japan}

\author{Kazuaki Kisu}%
\affiliation{Institute for Materials Research, Tohoku University, 2-1-1 Katahira, Aoba-ku, Sendai 980-8577, Japan.}

\author{Hao Li}%
\affiliation{Advanced Institute for Materials Research, Tohoku University, 2-1-1 Katahira, Aoba-ku, Sendai 980-8577, Japan}

\author{Shin-ichi Orimo}
\affiliation{Advanced Institute for Materials Research, Tohoku University, 2-1-1 Katahira, Aoba-ku, Sendai 980-8577, Japan}
\affiliation{Institute for Materials Research, Tohoku University, 2-1-1 Katahira, Aoba-ku, Sendai 980-8577, Japan.}
\email{ryuhei.sato.c1@tohoku.ac.jp (R.S.), shin-ichi.orimo.a6@tohoku.ac.jp (S.O.)}

\date{\today}

\begin{abstract}
 Topological data analysis based on persistent homology has been applied to the molecular dynamics simulation for the fast ion-conducting phase ($\alpha$-phase) of AgI, to show its effectiveness on the ion-migration mechanism analysis. Time-averaged persistence diagrams of $\alpha$-AgI, which quantitatively records the shape and size of the ring structures in the given atomic configurations, clearly showed the emergence of the four-membered rings formed by two Ag and two I ions at high temperatures. They were identified as common structures during the Ag ion migration. The averaged potential energy change due to the deformation of four-membered ring during Ag migration agrees well with the activation energy calculated from the conductivity Arrhenius plot. The concerted motion of two Ag ions via the four-membered ring was also successfully extracted from molecular dynamics simulations by our approach, providing the new insight into the specific mechanism of the concerted motion.
\end{abstract}

\maketitle

\section{\label{sec:level1}Introduction}
To realize all-solid-state ion batteries, ion-conducting electrolytes with high performance have been developed so far\cite{c4,c8,c5,c11,c6,c12,c1,c7,c2,c10,c3,c9}. These materials, called superionic conductors, tend to exhibit complex ion migration mechanisms, such as the ion migration with concerted motion\cite{c13,c14,c15,c16}, utilizing rotating anion\cite{c1,c2,c3,c4,c9,c10,c11,c12,c16,c17}, and the ion migration in glass materials\cite{c7}, which deviates from a single-ion hopping in a uniform lattice. Since atoms around mobile ions are arranged in a wide variety of configurations in such materials, extracting representative ion migration mechanisms and rate-limiting step is not easy. Even for bulk materials such as Li$_7$La$_3$Zr$_2$O$_{12}$, the activation energy for each Li ion migration during molecular dynamics (MD) simulations is widely distributed from 0 to 0.6 eV\cite{c18}. Therefore, it is challenging to construct a guideline for material design based on the migration mechanism.

Topological data analysis (TDA) based on persistent homology has been used for structural analysis of amorphous materials\cite{c19,c20,c21,c22}, glasses\cite{c23,c24,c26,c25,c53}, porous materials\cite{c27}, and complex molecular liquids\cite{c28}. This analysis can characterize point cloud data (ex. atomic structures), by projecting them onto a two-dimensional density distribution of the rings and pores in the system. Extracting representative ring/pore structures from the distribution and performing inverse analysis have contributed to the elucidation of the hidden order in the complex materials. In recent years, persistence diagrams, the distribution of ring and pore structures, have been combined with machine learning and other techniques to predict physical properties such as thermal conductivity properties\cite{c29,c30}, stress-strain curves\cite{c31}, formation enthalpy\cite{c32}, and so on\cite{c33}.

TDA was also used for dynamic phenomena. For example, it was applied for glass formation\cite{c25} and metal nucleation processes\cite{c34}. Based on defect chemistry, the ion migration is a phenomenon that occurs around ion vacancies. By considering vacancies in ionic conductors as large voids inside rings and pores in the persistence diagram, ion migration can be analyzed using TDA. By performing the inverse analysis along the ring/pore structures with high density, a representative ion migration mechanism within the material is considered to be extracted from the persistence diagram. Since the obtained rings/pores are composed of multiple ions, it is possible to discuss ion migration as a many-body problem.

In this study, we analyzed the Ag migration mechanism of $\alpha$-AgI\cite{c44,c43,c47,c38,c39,c50,c46,c45,c54}, a classical superionic conductor, as an example to elucidate complex migration mechanisms such as concerted motion. To our best knowledge, it is the first report to apply topological data analysis based on persistent homology to the ion migration study. The advantage of this method over existing ones is that it does not require any prior information like collective variables for constrained MD simulations and ion migration paths based on the crystallography, once the sufficient MD simulation trajectory is obtained. Note that the potential energy surface of the whole system mainly used in transition state search does not necessarily work for analyzing migration paths in these ionic conductors, because multiple ion migrations could occur simultaneously in a simulation cell. In addition, in the highly-disordered system, the activation energy obtained from nudged elastic band (NEB) method\cite{c51} is strongly affected by the position of mobile ions, because of the electrostatic interaction between them. This suggests that plenty of NEB calculations with different mobile-ion configurations are needed to obtain a reasonable ion migration mechanism. Therefore, this work becomes a milestone to understand the ion migration mechanism for highly-disordered system like glass materials\cite{c7} and compounds with complex-anions \cite{c1,c2,c3,c4,c9,c10,c11,c12,c16,c17}. 
\vspace{\fill}

\section{\label{sec:level2}Method}
\subsubsection*{A. Moleucular dynamics simulations}
NVT-MD simulations of $\alpha$-AgI were performed using MD calculation package (Large-scale Atomic/Molecular Massively Parallel Simulator (LAMMPS)\cite{c35} (version 29 October 2020)). Note that we chose $\alpha$-AgI as a model system to demonstrate the effectiveness of TDA for the ion migration mechanism for the following reasons; one is its high ionic conductivity\cite{c40,c43} making it easy to sample the ion migration during MD simulation. another is that the classical potential\cite{c36,c37} is available in this system, reducing the calculation cost. The other reason is that $\alpha$-AgI shows concerted motion of Ag ions\cite{c46,c50}, although it is consisted of ordered matrix, the bcc lattice of I ions. Therefore, it is easy to verify whether the ion migration mechanism obtained from TDA is reasonable or not, by comparing it with that suggested from the analysis and discussion based on the crystal structure. 

\begin{figure}
\includegraphics[width=\linewidth]{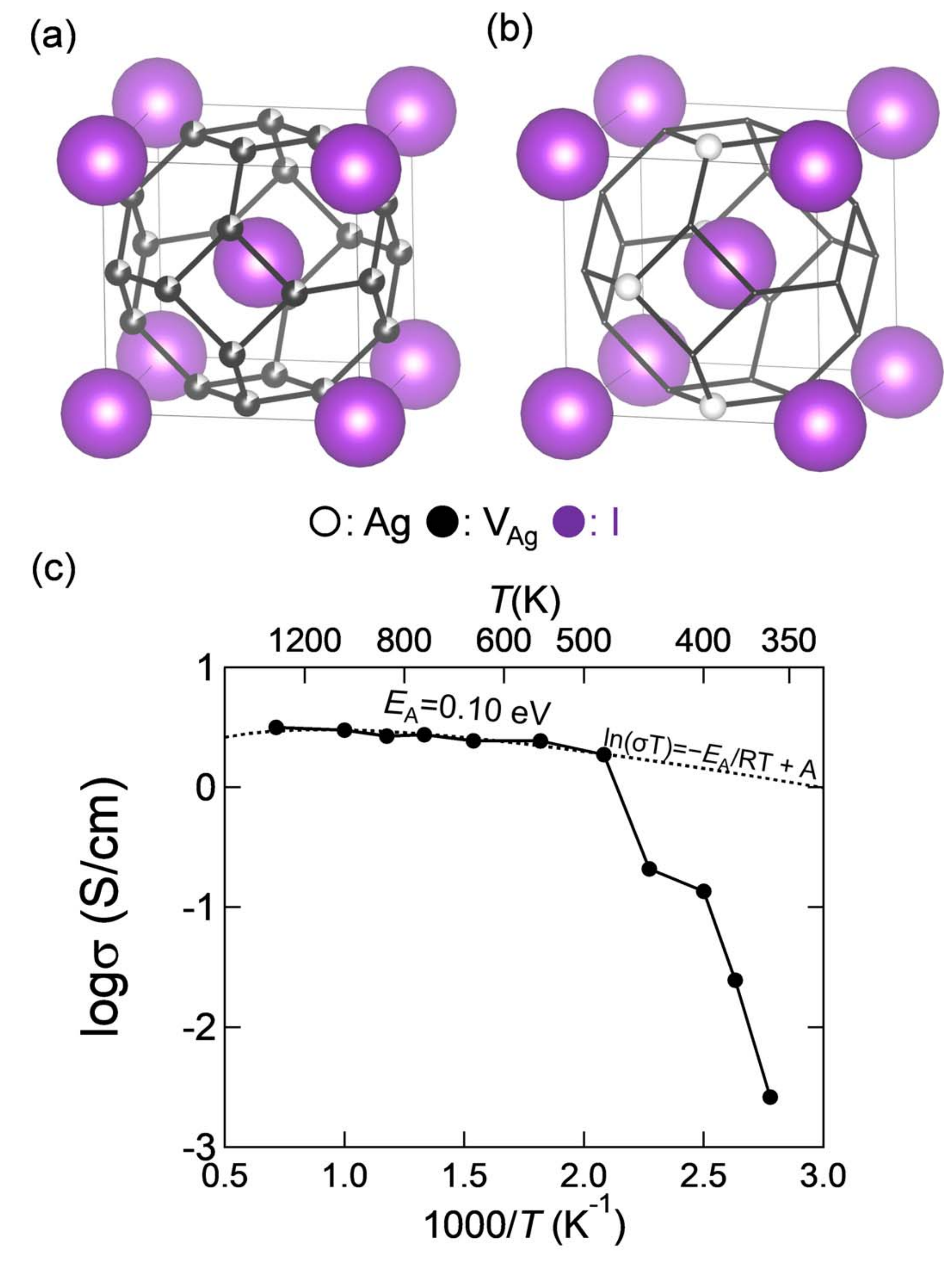}
\caption{\label{fig2}(a) Crystal structure of $\alpha$-AgI (white: Ag black: Ag vacancy, purple: I). I ions show a bcc structure, while Ag sites are located at the vertexes of the Voronoi polyhedron (black bold lines in the figure), which is composed of the perpendicular bisectors of each I ion. Ag ions occupy two of the twelve sites (12d sites). (b) An example of the structure of $\alpha$-AgI when Ag ions are specifically placed in two of these sites. (c) Conductivity Arrhenius plot of $\alpha$-AgI during classical MD simulations\cite{c42}. The dashed curve represents the fitting result of conductivity above 440K using Arrhenius equation, ln$\sigma T$ = --$E_A/RT$+A. 
}
\end{figure}

The detail of the calculation condition is followings. Nosé–Hoover thermostat and barostat\cite{c58} was employed for the temperature control in all MD simulations. Periodic boundary conditions were applied along all the Cartesian directions. Newton’s motion equations were integrated using the customary Verlet’s algorithm with a time-step length of 0.005 ps for MD simulations. 3-ns MD simulations with a cubic cell, 20.42 $\times$ 20.42 $\times$ 20.42 \AA, consisting of 256 Ag and I ions (4$\times$4$\times$4 supercell of $\alpha$-AgI (Fig. 1)) were performed at various temperatures from 360 to 1000K. Previous study \cite{c54} have reported that the finite size effect is negligibly small in AgI system, in terms of diffusion coefficients calculation. This suggests that our selected cell size does not affect the ion migration process during these MD simulations. We used the Parrinello-Rahman-Vashishta potential\cite{c36,c37}, for the interatomic potential for $\alpha$-AgI. It is known that Parrinello-Rahman-Vashishta potential represents Ag-I and I-I interactions well and can reproduce many physical properties such as diffusion coefficients\cite{c36} and phase transitions\cite{c37}. However, it has been reported that this potential cannot fully reproduce Ag-Ag interactions due to the lack of treatment for covalent interactions between Ag ions\cite{c38} and atomic polarization\cite{c39}. Nevertheless, the accuracy of the potential is not a matter of concern in this study, since our goal is to obtain the essential processes hidden in the complex ion migration dynamics containing concerted motions. Also note that NVT-MD simulations are intentionally conducted to maintain $\alpha$-AgI structure at low temperatures (360-440K). This makes us easier to compare the persistence diagram obtained at different temperatures since the matrix, I bcc lattice, and cell parameter is constant.  

To calculate time-averaged potential energy for ring-structures obtained from TDA, we also conducted constrained MD simulations. In these calculations, smaller cubic cell, 10.21 $\times$ 10.21 $\times$ 10.21 \AA, consisting of 32 Ag and I ions (2$\times$2$\times$2 supercell) was employed. In the simulation cell, we introduced one four-membered ring consisting of two Ag and two I ions obtained from TDA. Ions in this ring structure have been fixed during MD simulations by setting their force and velocity to zero. We prepared 16 simulation cells with different four-membered rings along the distribution obtained from the persistence diagram. Then, 4-ns MD simulations of these cells were conducted at 1000K. The difference of time-averaged potential energy between these simulations should be originated from the shape of four-membered ring and its surrounding environment, since the sampling time was long enough and ions except that in fixed four-membered ring could move freely. Therefore, we can discuss the change in potential energy during Ag ion migration via the four-membered ring based on time-averaged potential energy of these MD simulations.

\subsubsection*{B. Topological data analysis}
\begin{figure}
\includegraphics[width=\linewidth]{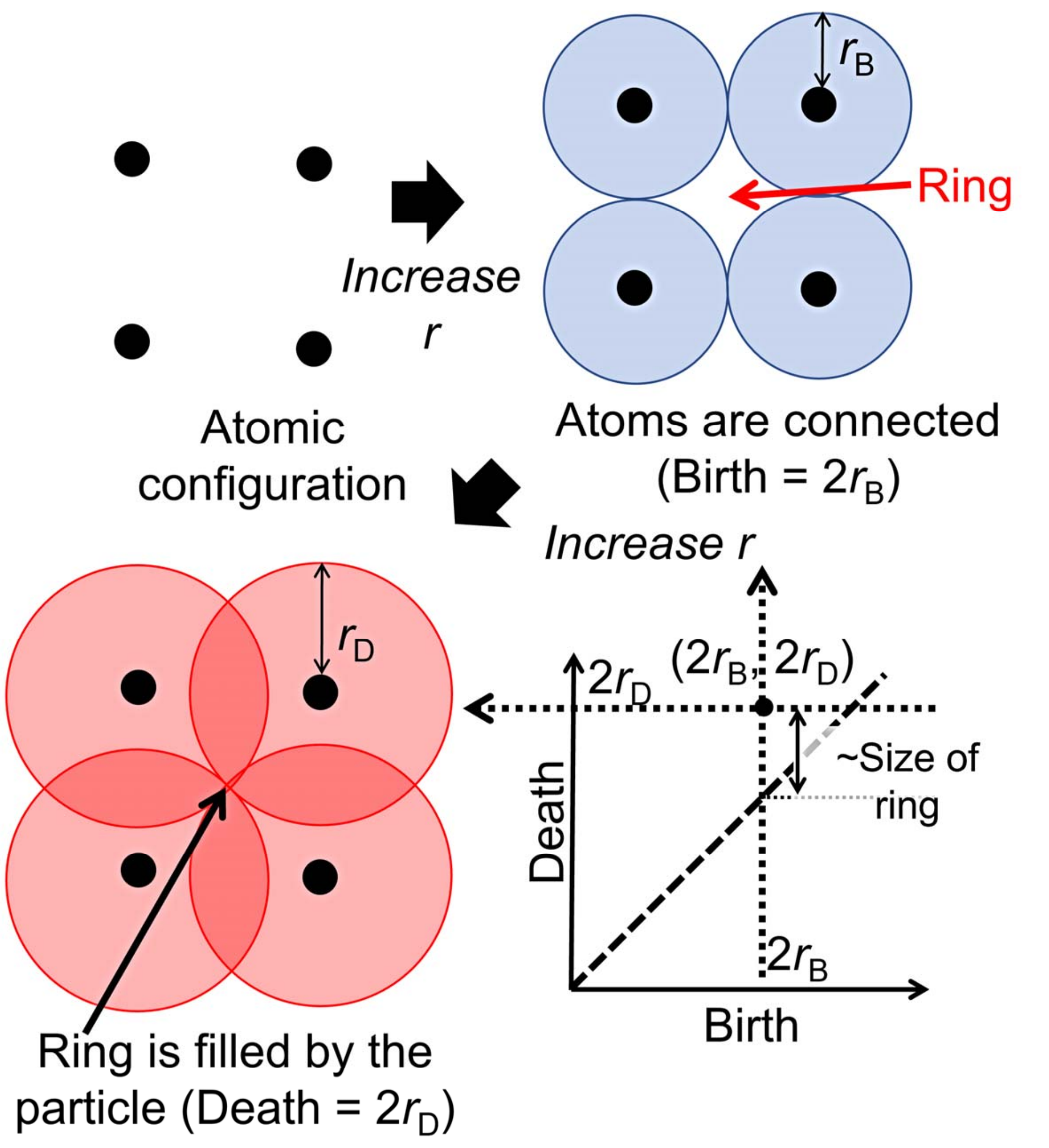}
\caption{\label{fig1} Schematic diagram of how ring structures in an atomic configuration are analyzed by topological data analysis (TDA) based on persistent homology. In this analysis, we place a sphere with same radius, $r$, at each atomic coordinate in the system. By increasing $r$, some atoms are connected to each other to form the ring. The diameter of the sphere when the ring is formed (2$r_b$) is assigned to the x-coordinate as birth diameter. At larger radius ($r = r_d$), this ring structure is filled by the spheres. Here, the diameter 2$r_d$ is named as death diameter and assigned to y-coordinate.}
\end{figure}

Persistent homology is a mathematical framework for analyzing n-dimensional holes in the given discrete data such as atomic coordinates. A software package, Homcloud\cite{c40,c41}, is employed for TDA based on persistent homology. Here, we only show a brief outline of its concept. See also reference 19 and 27 for the theoretical details and the application to materials. We will focus on one-dimensional holes (i.e. ring structures) hereafter. Note that one-dimensional hole in homology is defined as the states, which has a one-dimensional boundary (i.e., a space bounded by lines connecting atoms) but no two-dimensional boundary (i.e., the surface created by connecting atoms is not entirely filled by a sphere of radius $r$, assuming there is the sphere with a certain radius $r$ at each atomic coordinate). Figure 2 is a schematic diagram of how rings are analyzed in TDA based on persistent homology. First, a sphere with the common radius, $r$, is placed at each of atomic coordinates regardless of its element. By increasing $r$, new edges are spanned one by one when two spheres are overlapped. At a particular radius ($r = r_b$), some atoms are connected to each other and form a ring (top right figure in Fig. 2). In this paper, we name this diameter 2$r_b$ as the birth diameter and assign it to x-coordinate on the persistence diagram. At larger radius ($r = r_d$), this ring structure is filled by the sphere as shown in the bottom left figure. Here, the diameter 2$r_d$ is named as death diameter and assigned to y-coordinate. Thus, a persistence diagram is obtained as a two-dimensional scatter plot of birth-death pairs. To make it easy to understand visually, We limit the area to the square of [0.0, 12.0] $\times$ [0.0, 12.0] (\AA$^2$) and divided it into 200 $\times$ 200 meshes. We evaluated the number of birth-death pairs in each mesh as $log_{10}P(b,d)$.

\section{Results and discussion}
\subsubsection*{A. Change in the persistence diagram against ionic conductivity in $\bm{\alpha}$-AgI}

$\alpha$-AgI consists of a bcc I lattice with twelve stable Ag sites (12d sites), as shown in fig. 1(a). These stable Ag sites correspond to the vertex of the Voronoi polyhedron for the I bcc lattice, the polyhedron created by the perpendicular bisectors between I ions. Ag ions only occupy two of these stable sites like the structure in fig. 1(b). Therefore, once it is thermally activated, Ag ion can utilize unoccupied vacancy sites for ion migration. According to previous studies\cite{c43,c44}, six out of seven Ag ions migrate along the sides of the Voronoi polyhedron, the black line in the figure. On the other hand, the rest of Ag ions is transported through the face-center position (0.5 0.5 0) in the unit cell. Figure 1(c) is a conductivity Arrhenius plot of $\alpha$-AgI during MD simulations at various temperatures. We used Green-Kubo formula for the conductivity calculation\cite{c42,c55,c56,c57,c60,c61}. Due to the poor convergence of these calculations, the conductivity calculations used the results of MD simulations extended to 25 ns. $\alpha$-AgI showed superionic conduction at temperatures higher than 440 K. The obtained conductivity above 440 K is in good agreement with the experimental value ($\approx10^0$ (S/cm))\cite{c43}.

\begin{figure}
\includegraphics[width=\linewidth]{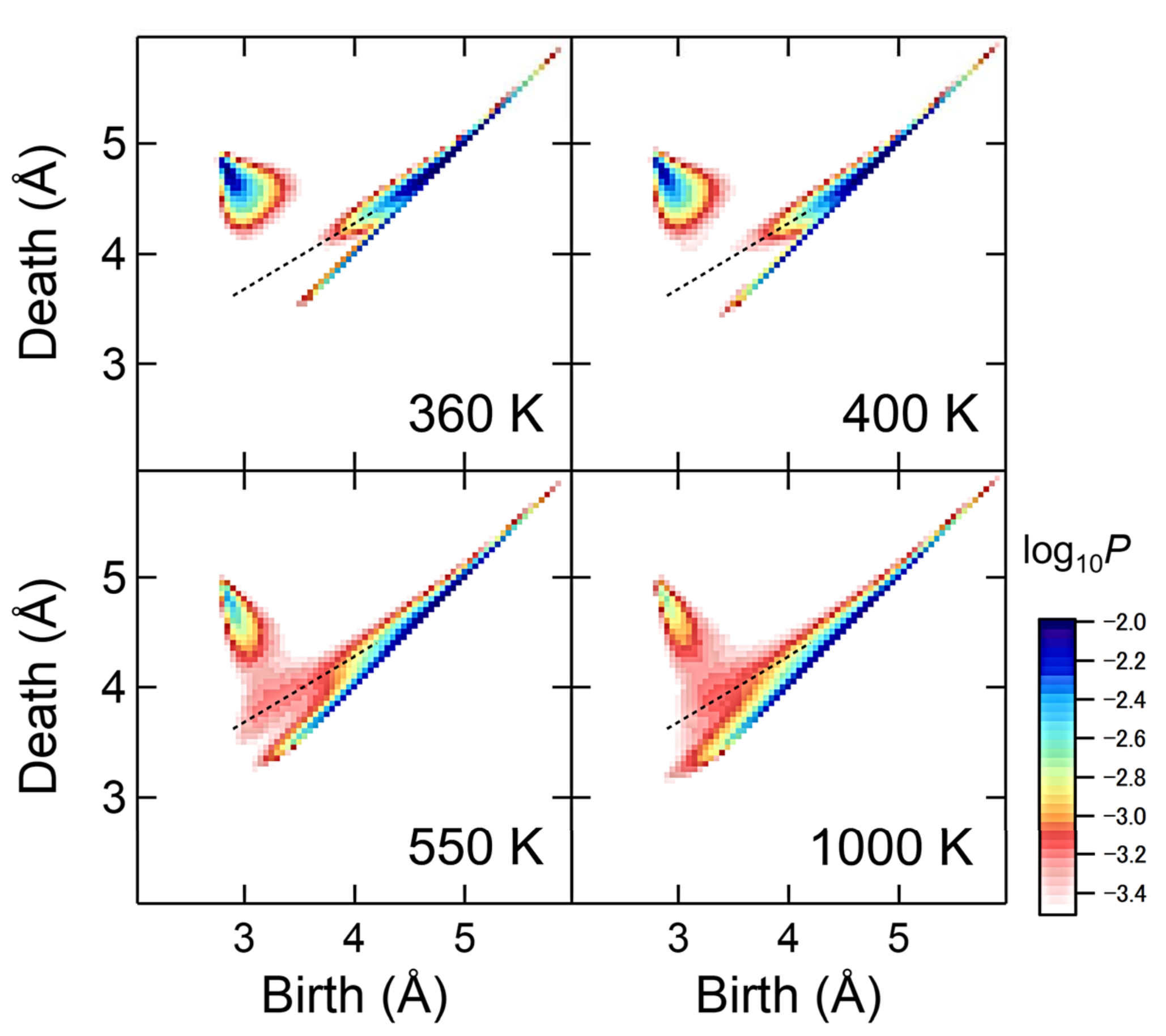}
\caption{\label{fig3} Time-averaged persistence diagram of the ring structure during MD simulations of $\alpha$-AgI for several temperatures. This persistence diagram represents a probability density ($P$) distribution normalized so that the integrated value over the squared regions [0.0, 12.0] $\times$ [0.0, 12.0] (\AA$^2$) becomes unity. The color in the figure is scaled according to log$_{10}P$ and represents the probability density per the mesh of 0.05 $\times$ 0.05 \AA$^2$.}
\end{figure} 

The change in the local structure of $\alpha$-AgI during Ag migration has been confirmed prominently as that in ring structures in the persistence diagram. Figure 3 shows time-averaged persistence diagrams during MD simulations at various temperatures. We conducted TDA for the atomic configurations every 0.05 ps during MD simulations and obtained the averaged persistence diagram of 6,000 configurations. Two types of ring structures are mainly distributed around (Birth, Death)=(b,d)=(2.8,4.5) and (4.2,4.4) in the persistence diagrams for MD simulations at 360K and 400K, where ionic conductivity is low. On the other hand, the ring-structure distribution along the black dashed line increases on the persistence diagrams at 550K and 1000K. Therefore, these ring structures could correspond to the ones for Ag migration. Note that the same change in the ring structure has been confirmed in the persistent diagrams obtained from the NPT-MD simulations, taking into account the phase transition to $\beta$-AgI (see section S1 in the supporting information).

\subsubsection*{B. Inverse analysis for the persistence diagram of $\bm{\alpha}$-AgI}

\begin{figure}
\includegraphics[width=\linewidth]{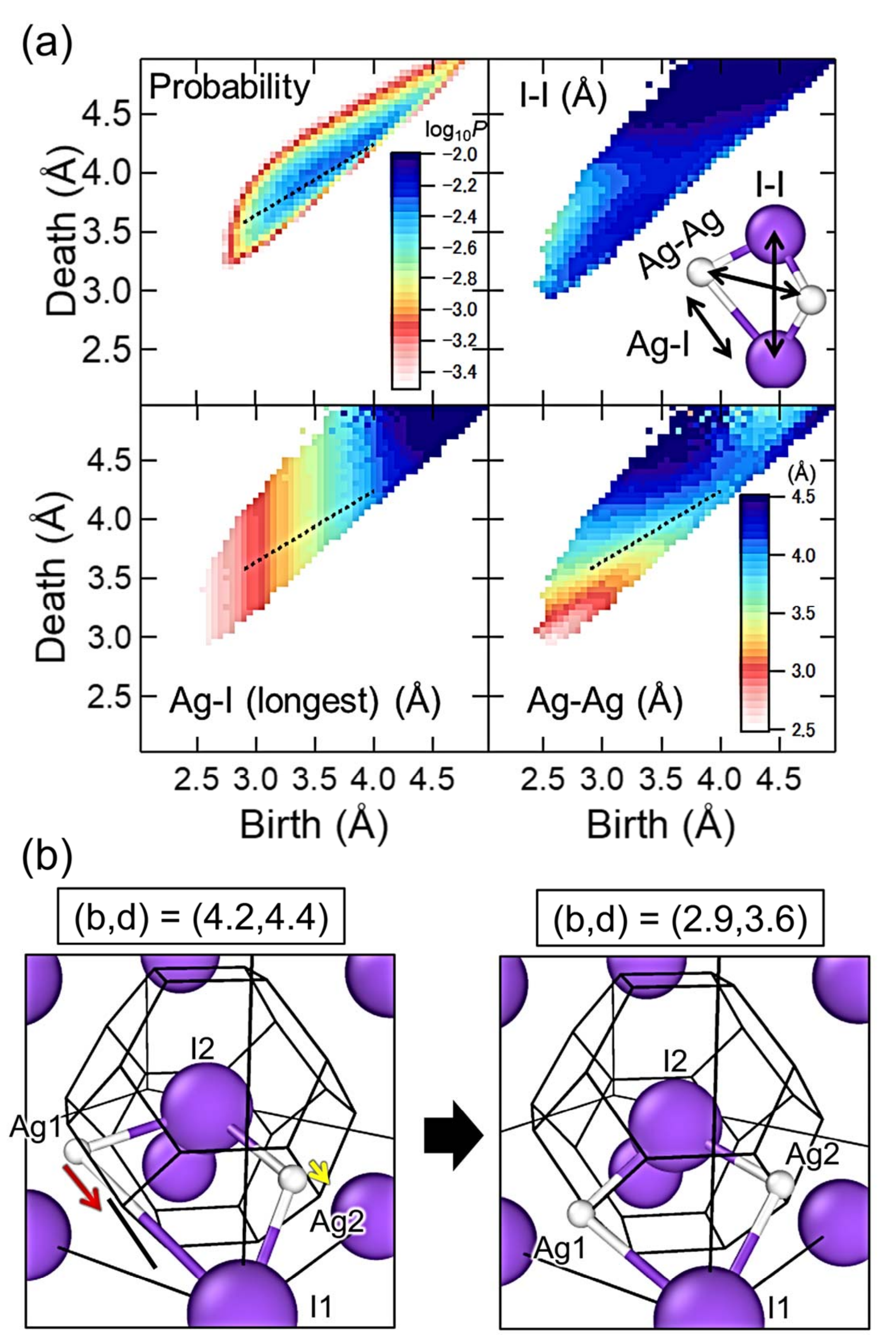}
\caption{\label{fig4} (a) Persistence diagram only showing four-membered rings and the distribution of averaged I-I, Ag-I, and Ag-Ag distances in each (b,d) for these four-membered rings in the persistence diagram. (Calculated for each 0.05 $\times$ 0.05 \AA$^2$ meshes.) The color in top left figure is scaled according to log$_{10}P$ and represents the probability density per the mesh of 0.05 $\times$ 0.05 \AA$^2$. On the other hand, that in other figures is scaled according to the distance (\AA) between ions. (b) Ag migration obtained by the deformation of four-membered ring along the black dashed line on the persistence diagram. Note that I ions except that in the four-membered ring show ideal I bcc lattice positions. (white: Ag, purple: I)}
\end{figure}

From the inverse analysis\cite{c41} of the persistence diagram, a model of Ag ion migration via four-membered rings has been obtained. Here, the inverse analysis means to calculate averaged property for each mesh in the persistence diagram, recorded during the TDA. Inverse analysis of the persistence diagram at 1000K confirmed that these ring structures form a four-membered ring consisting of two Ag and I ions (Fig. S2). The persistence diagram of the four-membered rings in the system and the results of their inverse analysis (the average I-I, Ag-I, and Ag-Ag distances in the four-membered rings at each mesh) are summarized in Fig. 4(a). As shown in the figure, the distance between I ions is almost constant. On the other hand, the length of longest Ag-I bond in the four-membered ring decreases as birth decreases. Similarly, the distance between Ag ions was also shortening with decreasing death. These results show that the four-membered ring deforms, as the birth decreases along the dashed black line in the persistence diagram. Figure 4(b) shows the structure of the four-membered ring reproduced based on these average distances. Note that I ions other than that in the 4-membered ring occupy the ideal positions of bcc structure in Fig. 4(b). The black line is the Voronoi polyhedron of an ideal I bcc lattice ($\approx$Ag migration path). Ag1 moved from one vertex to the other in the Voronoi polyhedron, indicating that Ag hopping migration has been reproduced by the deformation of the four-membered ring obtained from the persistence diagram.

In these four-membered rings, the interaction between Ag1 and Ag2 ions via the I-lattice has been confirmed from bond valence analysis (Fig. S3). Such a picture of Ag migration via interaction with I-lattice has been claimed in previous studies. For example, covalent-bond formation between nearest neighbour Ag and I ions has been reported from the bond-overlap population analysis of $ab$ $initio$ MD simulation for molten AgI\cite{c38}. In addition, $ab$ $initio$ MD simulation and THz-frequency Raman polarization-orientation measurement suggest that anharmonic vibrations in the I lattice are important for Ag migration\cite{c45}.

\begin{figure}
\includegraphics[width=\linewidth]{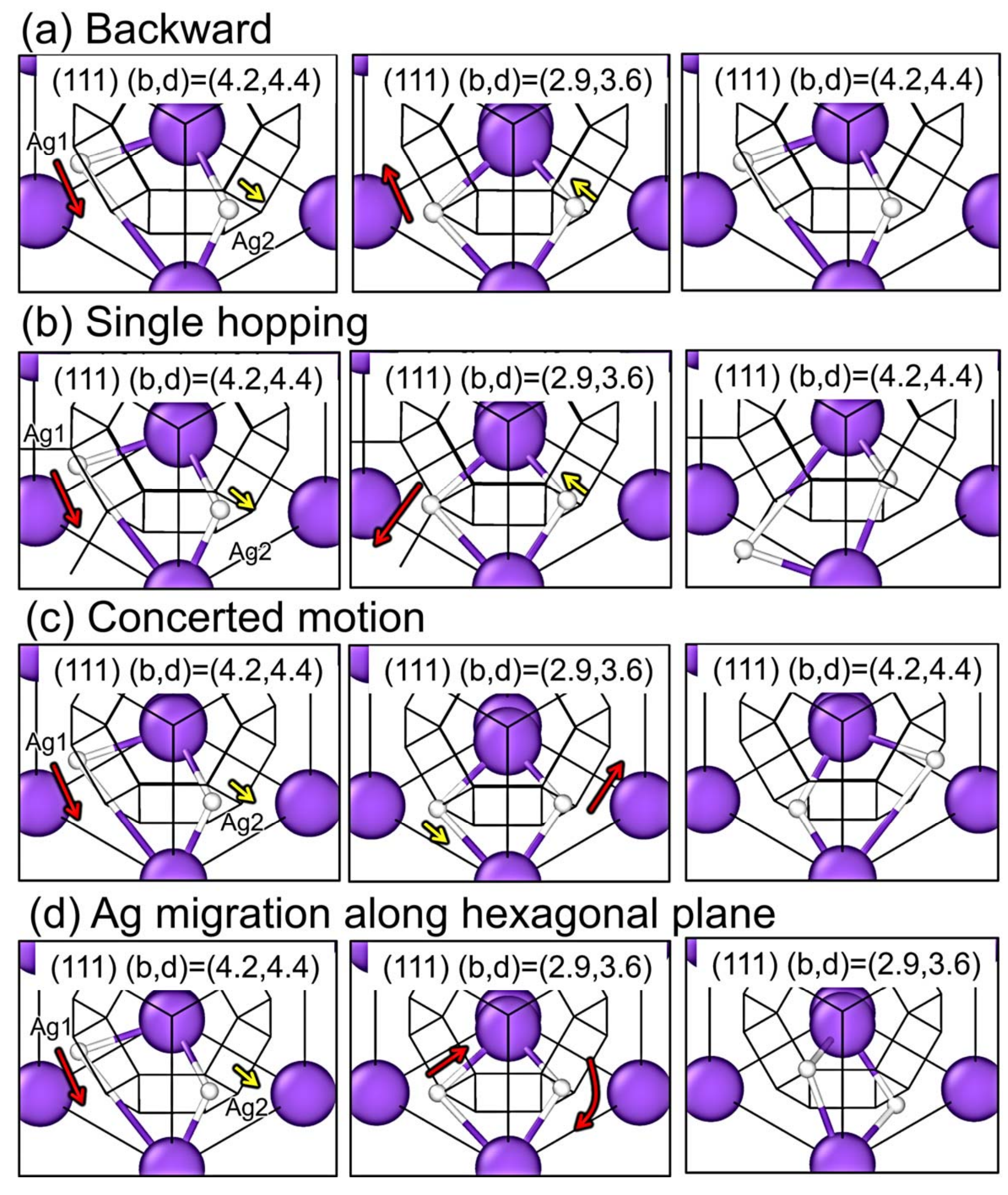}
\caption{\label{fig5} Ag migration obtained by repeating the deformation of the four-membered ring as shown in Fig. 4(b) (i.e., high (b,d) $\rightarrow$ low (b,d) $\rightarrow$ high (b,d)); (a) backward Ag migration, (b) single Ag hopping, (c) concerted motion of two Ag ions, and (d) concerted motion of Ag ions maintaining the four-membered ring shape. The figure is created with (111) as the normal vector, and (b,d) represents the values of birth and death of the four-member ring. (white: Ag, purple: I)}
\end{figure}

Fig. 5 and the video in Supporting information summarize possible Ag migrations via the four-membered ring. During Ag migration in figs. 5(a)-(c), the four-membered ring deforms along the dashed black line in Fig. 4(a). On the other hand, the four-membered ring shape is maintained in the latter part of Ag migration in Fig. 5(d). There are four possible migrations. The first one (Fig. 5(a)) shows Ag hopping and subsequent returning to its original site. In the second one (Fig. 5(b)), Ag1 further migrates to the other site. In this reaction, Ag2 returns to the almost same position. Therefore, this migration is considered to be the single Ag1 hopping. The third one (Fig. 5(c)) is a concerted motion, where Ag1 migrates from one site to the other and Ag2 hopping occurs immediately after Ag1 migration. Fig. 5(d) also shows a concerted migration of Ag ions. However, it occurs without the deformation of the four-membered ring. As shown in the figure, Ag ions migrate along the hexagon on the Voronoi polyhedron of I bcc, maintaining the shape of the four-membered ring.

From the analysis of Ag hopping directions between nearest neighbour sites, the ratio of these four Ag migrations are estimated to be around (a) 42\%, (b) 19\%, (c) 26\%, and (d) 13\%, respectively (Fig. S4). This suggests that concerted motions such as (c) and (d) occur with about 39\% probability. The existence of concerted motion in AgI and similar class of material (Ag$_2$Se) has been claimed based on the Haven ratio for ionic conductivity\cite{c46} and the velocity correlation function of Ag ions during MD simulations\cite{c47}. These results are consistent with the obtained concerted motion in Fig. 5(c) and (d). The advantage of TDA from conventional studies is that we can extract the concrete model of concerted ion migration as shown in Fig. 5. 

\begin{figure}
\includegraphics[width=\linewidth]{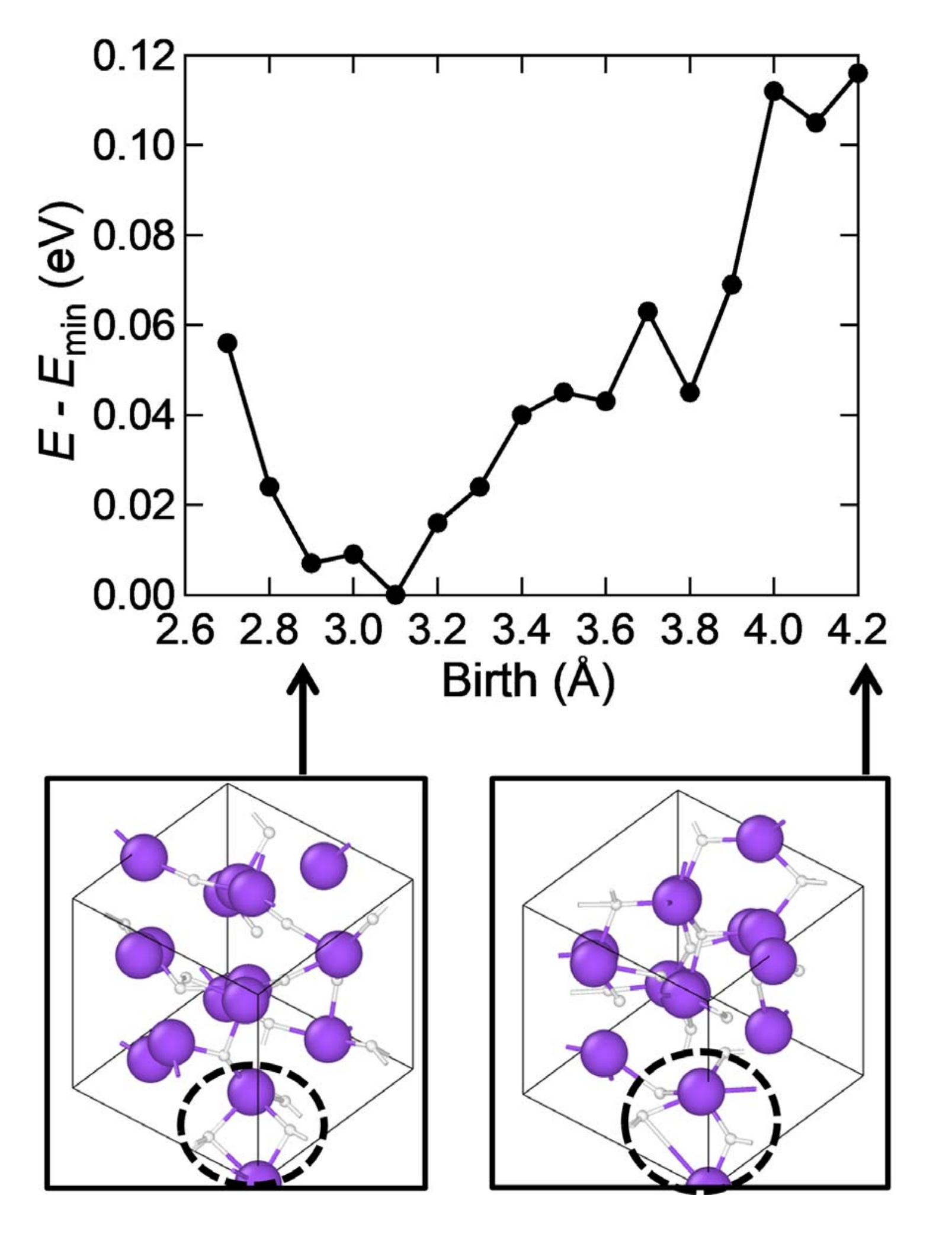}
\caption{\label{fig6} Time-averaged energy during 4-ns MD simulations for 2$\times$2$\times$2 $\alpha$-AgI supercells with one fixed four-membered ring obtained from a persistence diagram along the black dashed line in Fig. 4(a).}
\end{figure}

The Ag migration via the four-membered ring obtained from the persistence diagram corresponds to the representative Ag migration in $\alpha$-AgI. Figure 6 shows the time-averaged potential energy during 4-ns MD simulations with four-membered rings obtained from the persistence diagram fixed in a 2×2×2 $\alpha$-AgI supercell. Since ions other than that in the four-membered ring moved freely and the sampling time was long enough, the energy difference in these MD simulations originated from the four-membered ring shape. Therefore, fig. 6 shows the energy profile for the deformation of the four-membered ring along the black dashed line in the persistence diagram in Fig. 4(a). The averaged potential energy change due to the deformation of four-membered ring during Ag migration is about 0.12 eV. Note that the energy profile for three types of Ag migration in fig. 5(a)-(c) should be the same as that of Fig. 6. This is because these Ag migrations can be reproduced by a point-symmetric operation of the four-membered ring on the black dashed line in Fig. 4(a) using the middle point of two I ions in the ring. The averaged potential energy change due to the deformation of four-membered ring, which can reproduce Ag migration, is in good agreement with the activation energy obtained from the conductivity Arrhenius plot. This shows that the persistence diagram is able to extract a local but representative structure for Ag migration.

\subsubsection*{C. The comparison between independent and  four-membered ring Ag migration}
\begin{figure}
\includegraphics[width=\linewidth]{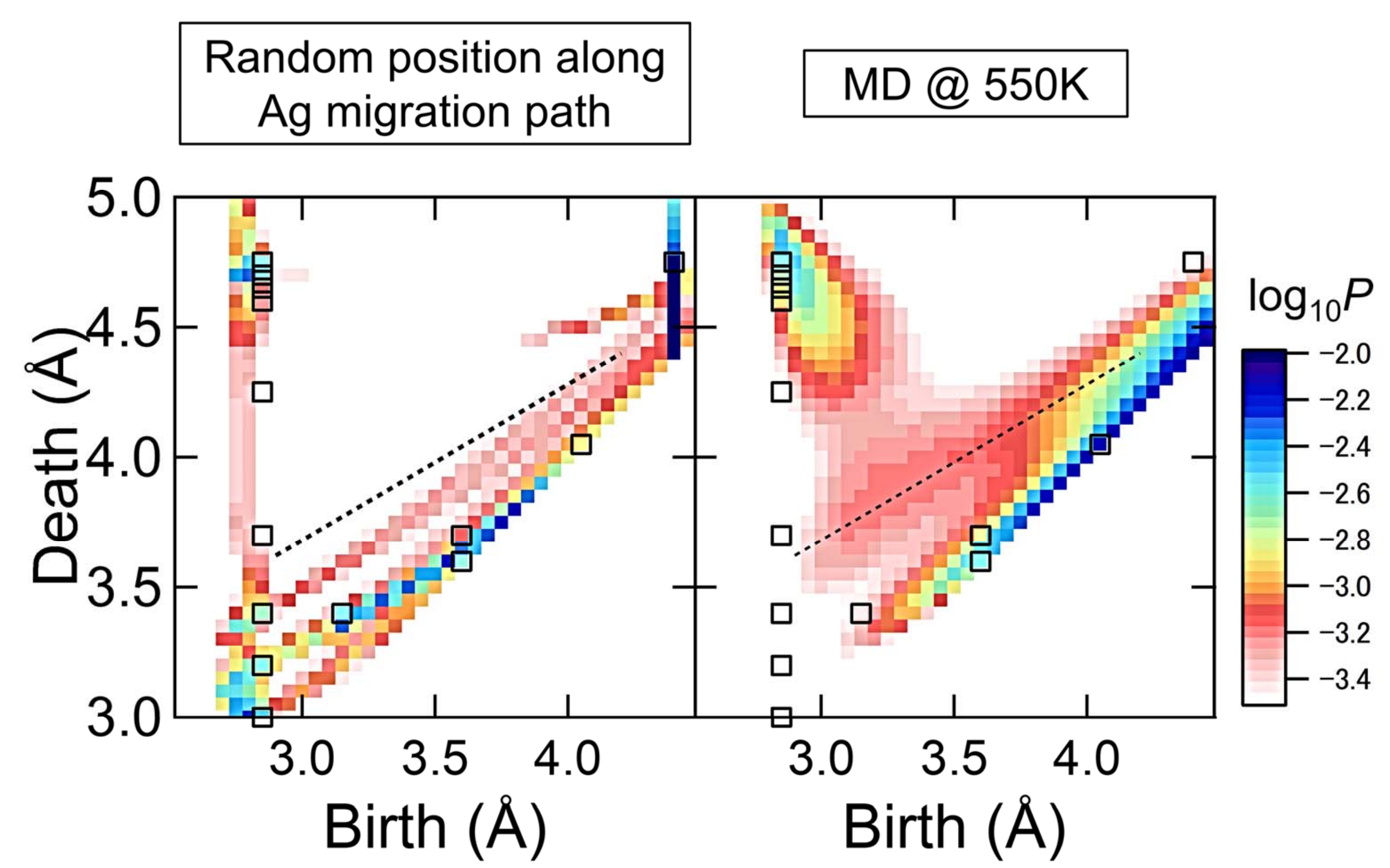}
\caption{\label{fig7} (left) Persistence diagram of I bcc lattice with Ag ions randomly placed on the migration path, (edges and vertexes of the Voronoi polyhedron in fig. 1(a)). (right) Time-averaged persistence diagram during MD simulation of $\alpha$-AgI at 550 K. The black open squares in the figure represent the (b,d) positions of the ring structure obtained from I bcc lattice with Ag ions randomly placed at the stable (12d) site (edges of the Voronoi polyhedron).The color in the figure is scaled according to log$_{10}P$ and represents the probability density per the mesh of 0.05 $\times$ 0.05 \AA$^2$.}
\end{figure}

Independent migration of each Ag ion like a random-walk model\cite{c49} does not form the four-membered rings specific to the reproduced Ag migration, showing that the interaction between ions selectively creates them during MD simulations. Figure 7(left) shows a persistence diagram of an ideal I bcc lattice with Ag ions randomly placed on the migration path (i.e. the edges and vertexes of the Voronoi polyhedron for I bcc lattice). Since Ag ions are placed almost randomly along the migration path\cite{c48}, the resulting persistence diagram is considered to be equivalent to that for the random-walk model. We created 6000 random configurations of 4$\times$4$\times$4 supercells and calculated their averaged persistence diagram so that the calculation condition for it matches that for the persistence diagram for MD simulations. The black open squares in the figure represent the (b,d) positions of rings obtained when Ag ions are only placed at the stable site (the vertexes of the Voronoi polyhedron). The distribution among these black open squares was found to increase, when Ag ions are randomly placed on the migration path in the I bcc lattice. However, ring structures along the black dashed line observed in MD simulations at 550 and 1000K were hardly obtained in the persistence diagram of I bcc lattice with random Ag configurations. Therefore, we conclude that independent migration of Ag ions cannot explain the ones during MD simulations and the concerted motion of Ag ions induced by the interaction between them actually contributes to Ag ion conductivity in $\alpha$-AgI.

\section{Conclusion}
To reveal the ion migration mechanism in the superionic conductor, we performed classical molecular dynamics simulations for $\alpha$-AgI and made a series of topological data analysis based on persistent homology. When $\alpha$-AgI shows superionic conductivity, characteristic Ag-I ring structures were newly detected on the persistence diagram. They were identified as four-membered rings formed by two Ag and I ions, which are common to the Ag ion migration. The averaged potential energy change due to the deformation of four-membered ring during Ag migration is about 0.12 eV, which is in good agreement with the activation energy obtained from the conductivity Arrhenius plot. Therefore, we conclude that the Ag ion migration via these four-membered rings is the representative Ag migration process in $\alpha$-AgI. In this migration, two Ag ions showed concerted motion and the ratio of such Ag ion migration was about 39\%. On the other hand, the number of the corresponding four-membered rings became significantly small, if random-walk of Ag ions in I bcc lattice is assumed. Therefore, based on topological data analysis of the MD trajectories, we conclude that it is successfully elucidated that Ag ions migrate not independently but concertedly via the four-membered ring in $\alpha$-AgI. 

\section*{SUPPLEMENTARY MATERIAL}
See the supplementary material for the persistence diagram of AgI during 3-ns NPT-MD simulations (section S1), the distribution of most frequent number of atoms in the ring in persistence diagram during the MD simulation at 1000K (section S2), bond valence analysis of four-membered rings obtained from inverse analysis of persistence diagram (section S3), The result of the direction analysis of single Ag hopping during MD simulation at 1000K (section S4), and videos for Ag migrations through the deformation of four-membered ring obtained from the topological data analysis based on persistent homology.

\begin{acknowledgments}
This work was supported by JSPS KAKENHI Grant-in-Aid for Scientific Research on Innovative Areas “Hydrogenomics”, No. JP18H05513 and Grant-in-Aid for Challenging Research (Pioneering, No, 19H05514). The computation in this work has been done using the facilities of the Supercomputer Center, the Institute for Solid State Physics, the University of Tokyo.
\end{acknowledgments}

\section*{AUTHOR DECLARATIONS}
\section*{Conflict of Interest}
The authors have no conflicts to disclose.
\section*{Data Availability}
The data that support the findings of this study are available from the corresponding author upon reasonable request. Note that the input files for MD simulations, representative output files, and the calculation codes used in this paper are available on Materials Cloud platform\cite{c59}. See DOI: https://doi.org/10.24435/materialscloud:zn-5t 

\section*{Reference}
\nocite{*}
\bibliography{aipsamp}

\begin{thebibliography}{60}
\bibitem [{c13()}]{c13}%
  \BibitemOpen
  \href@noop {} {\ }\bibinfo {note} {The migration of multiple ions triggered
  by that of one ion. Here, the term, concerted motion, is defined in a broad
  sense. In this article, if other ions move in conjunction with the migration
  of one ion, it is considered to be a concerted motion, regardless of the
  direction of transport, etc.}\BibitemShut {Stop}%
\bibitem [{c42()}]{c42}%
  \BibitemOpen
  \href@noop {} {\ }\bibinfo {note} {The following Green-Kubo formula was
  employed to calculate Ag-ion conductivity, \begin{equation}
  \sigma=\frac{1}{3Vk_bT} \int_0^{\infty}
  \langle\sum_iq_i\bm{v_i}(t')\cdot\sum_iq_i\bm{v_i}(t'+t)\rangle_{t'} dt
  \rm{,} \end{equation} where $v_i$ and $q_i$ are is velocity and the formal
  charge of ith ion. $V$ is the volume of the calculation cell. $t$ and $t'$
  are MD time step, and MD time step for the reference velocity,
  respectively.The inside of the bracket represents the correlation function of
  the current for ions.}\BibitemShut {Stop}%
\bibitem [{c49()}]{c49}%
  \BibitemOpen
  \href@noop {} {\ }\bibinfo {note} {Here, we presume that each Ag ion migrates
  independently along the edge of the Voronoi polyhedron for I bcc lattice
  without interacting each others.}\BibitemShut {Stop}%
\end{thebibliography}

\end{document}